\documentclass[aip,cha,amsmath,amssymb,reprint,author-year,author-numerical]{revtex4-1}
\usepackage{epsfig}
\usepackage{color}
\usepackage{graphicx}
\usepackage{dcolumn}
\usepackage{bm}
\usepackage[shortlabels]{enumitem}
\input epsf

\usepackage[normalem]{ulem} 

\graphicspath{{figures/}}

\begin{document}


\title{Multistability in coupled oscillator systems with higher-order interactions and community structure}

\author{Per Sebastian Skardal}
\email{persebastian.skardal@trincoll.edu} 
\affiliation{Department of Mathematics, Trinity College, Hartford, CT 06106, USA}

\author{Sabina Adhikari}
\affiliation{Department of Applied Mathematics, University of Colorado at Boulder, Boulder, Colorado 80309, USA}

\author{Juan G. Restrepo}
\affiliation{Department of Applied Mathematics, University of Colorado at Boulder, Boulder, Colorado 80309, USA}

\begin{abstract}
We study synchronization dynamics in populations of coupled phase oscillators with higher-order interactions and community structure. We find that the combination of these two properties gives rise to a number of states unsupported by either higher-order interactions or community structure alone, including synchronized states with communities organized into clusters in-phase, anti-phase, and a novel skew-phase, as well as an incoherent-synchronized state. Moreover, the system displays a strong multistability, with many of these states stable at the same time. We demonstrate our findings by deriving the low dimensional dynamics of the system and examining the system's bifurcations using a stability analysis and perturbation theory.
\end{abstract}


\maketitle

\begin{quotation}
Spontaneous entrainment and pattern formation in large ensembles of coupled oscillator units play a role in a wide range of applications in mathematics, physics, engineering, and biology~\cite{Strogatz2003,Pikovsky2003}. Recent insights from both neuroscience and physics~\cite{Petri2014Interface,Ashwin2016PhysD} point to the presence of higher-order interactions in populations of coupled dynamical systems. While initial studies have begun to uncover the effects that higher-order interactions have on macroscopic system dynamics, the role that both small- and large-scale structural properties of networks play in shaping collective dynamics in the presence of higher-order interactions remains unexplored and poorly understood. Here we study the dynamics of systems of coupled phase oscillators with community structure and higher-order interactions. Using dimensionality reduction methods to derive the macroscopic system dynamics, we uncover and analytically characterize a range of new, coexisting states that are not supported by higher-order interactions or community structure alone.
\end{quotation}

\section{Introduction}\label{sec1}

Synchronization of coupled oscillators is one of the most prominent examples of emergent collective behavior, with examples including circadian rhythms~\cite{Lu2016Chaos}, brain rhythms~\cite{KitzbichlerPLOS2009}, cardiac excitation~\cite{Glass1988}, pedestrian synchronization in suspension bridges~\cite{Strogatz2005Nature}, Josephson junctions~\cite{Wiesenfeld1996PRL}, and power grids~\cite{Rohden2012PRL}. The Kuramoto model~\cite{Kuramoto1984} of coupled phase oscillators has been particularly useful for shedding light on synchronization phenomena due to its analytic tractability and versatility for modeling many different systems with various physical properties. While the Kuramoto model describes the pairwise interactions of oscillators, it does not take into account the possibility of interactions involving multiple nodes simultaneously, i.e., higher-order interactions. In recent years, the role of higher-order interactions in shaping the collective response of coupled dynamical systems has received much interest~\cite{Battiston2020PhysRep,Majhi2022Interface}. Higher-order interactions have been shown to qualitatively change the dynamics of epidemic and opinion dynamics~\cite{Iacopini2019NatComms,Matamalas2019}, the behavior of ecological systems~\cite{GrilliNature2017}, and other systems that display emergent collective dynamics~\cite{Carletti2020JPhys}. For coupled oscillator systems, higher-order interactions are motivated by applications in neuroscience~\cite{Petri2014Interface,Giusti2016JCN,Reimann2017,Sizemore2018JCN} and physics~\cite{Ashwin2016PhysD,Leon2019PRE}. Higher-order interactions can cause dramatic changes to the dynamics of coupled oscillators, including bistability, explosive transitions, and extensive multistability \cite{Skardal2019PRL,Skardal2020CommPhys,Xu2020PRR,Skardal2020JPhys,Xu2021PRR,Skardal2021PRR,Anwar2022Chaos,Skardal2022Chaos,Leon2022PRE,Kachhvah2022NJP,Kundu2022PRE}. The Kuramoto model has also been generalized to include phases defined not only on the nodes of a network, but on each edge defining a higher-order interaction \cite{Millan2020PRL,Ghorbanchian2021Communications,DeVille2021Chaos} (a hyperedge when the interactions are encoded in a hypergraph, or a simplex when they are encoded in a simplicial complex).

For the network Kuramoto model with pairwise interactions, the role of network community structure is well studied, as it gives rise to a variety of phenomena including hierarchical synchronization~\cite{Skardal2012PRE}, chimera states~\cite{Abram2008PRL}, and nonmonotonic synchronization transitions~\cite{Restrepo2019PRR}. Despite the prevalence of community structure in a wide range of real-world networks~\cite{Girvan2002PNAS,Newman2006PNAS}, the effect of community structure in networks with higher-order interactions is not yet studied. In this paper we study a generalization of the Kuramoto model that includes higher-order interactions and community structure. We apply the Ott-Antonsen ansatz~\cite{Ott2008Chaos,Ott2009Chaos} to obtain a low-dimensional system of equations. Focusing on the case of two communities, we find that, depending on the strength of the community structure, the system admits multiple coexisting stable synchronization states. In addition to both communities being incoherent and both communities being synchronized and in-phase synchronized with one another, i.e., entrained with a macroscopic phase difference $\phi=0$, a handful of more complicated states exist. In one such state one community is synchronized while the other is (nearly) incoherent. Another state is characterized by both communities being synchronized, but each community is anti-phase synchronized with one another, i.e., forming two clusters with a phase difference $\phi=\pi$. Finally, when inter-community coupling is negative a surprising state emerges where both communities are synchronized, but are neither in-phase nor anti-phase, rather the two communities form two clusters with a phase difference $0<\phi<\pi/2$.
These states might find relevance in systems of oscillators where higher-order interactions are present, such as brain rhythms.

The paper is organized as follows. In Sec.~\ref{sec:02} we present our model and relevant notation. In Sec.~\ref{sec:03} we derive a low-dimensional description of the dynamics by using the Ott-Antonsen ansatz. In Sec.~\ref{sec:04} we study in detail the case of two communities by analyzing particular cases and validate our theoretical analysis with numerical results. In Sec.~\ref{sec:05} we present our conclusions.

\section{Governing equations}\label{sec:02}

We consider the dynamics of a large ensemble of $N$ phase oscillators organized into $C$ communities. In general we assume that each community $\sigma$ consists of $N_\sigma$ oscillators, with $\sigma=1,\dots,C$ indexing each community, so that $\sum_{\sigma=1}^CN_\sigma=N$. For simplicity, we consider both dyadic and triadic (i.e., pair-wise and triplet) interactions, although our results could be extended to interactions involving more nodes. Following Ref.~\cite{Skardal2020CommPhys}, the dynamics of each oscillator is given by
\begin{align}
\dot{\theta}_i^\sigma&=\omega_i^\sigma+\sum_{\sigma'=1}^C\frac{K_1^{\sigma\sigma'}}{N_{\sigma'}}\sum_{j=1}^{N_{\sigma'}}\sin(\theta_j^{\sigma'}-\theta_i^{\sigma})\nonumber\\
&+\sum_{\sigma'=1}^C\sum_{\sigma''=1}^C\frac{K_2^{\sigma\sigma'\sigma''}}{N_{\sigma'}N_{\sigma''}}\sum_{j=1}^{N_{\sigma'}}\sum_{l=1}^{N_{\sigma''}}\sin(2\theta_j^{\sigma'}-\theta_l^{\sigma''}-\theta_i^{\sigma}),\label{eq:01}
\end{align}
where $\theta_i^\sigma$ and $\omega_i^\sigma$ denote the phase and natural frequency of oscillator $i$ in community $\sigma$, $K_1^{\sigma\sigma'}$ is the dyadic coupling strength between a pair of oscillators in communities $\sigma$ and $\sigma'$, respectively, and $K_2^{\sigma\sigma'\sigma''}$ is the tetradic coupling strength between a triplet of oscillators in communities $\sigma$, $\sigma'$, and $\sigma''$, respectively. We emphasize that this all-to-all topology with community-wise uniformity in coupling eliminates the complications of both intra- and inter-community structures that may blur the fundamental effects of community structure. We will assume that the natural frequencies of oscillators in each community $\sigma$ are drawn from the distribution $g^\sigma(\omega)$, which we discuss later.

Next, we introduce the generalized community-wise order parameters
\begin{align}
z_q^{\sigma}=r_q^{\sigma}e^{i\psi_q^\sigma}=\sum_{j=1}^{N_\sigma}e^{qi\theta_j^\sigma}.\label{eq:02}
\end{align}
Note that for the case of $q=1$ we recover the classical Kuramoto order parameter $z^\sigma=r^\sigma e^{i\psi^\sigma}$ for each community, whose magnitude $r^\sigma$ quantifies the degree of synchronization in community $\sigma$, while higher-order modes, e.g., $q=2$, measure cluster synchronization~\cite{Skardal2011PRE}. Using the generalized order parameters in Eq.~(\ref{eq:02}) for $q=1$ and $2$ we may rewrite Eq.~(\ref{eq:01}) as
\begin{align}
\dot{\theta}_i^\sigma=\omega_i^\sigma + \frac{1}{2i}\left(He^{-i\theta_i^\sigma}-H^*e^{i\theta_i^\sigma}\right),\label{eq:03}
\end{align}
where
\begin{align}
H=\left(\sum_{\sigma'=1}^C K_1^{\sigma\sigma'}z^{\sigma'}\right) + \left(\sum_{\sigma'=1}^C\sum_{\sigma''=1}^C K_2^{\sigma\sigma'\sigma''}z_2^{\sigma'}z^{\sigma''*}\right),\label{eq:04}
\end{align}
where $*$ denotes the complex conjugate and we let $z^\sigma=z_1^\sigma$.

\section{Reduced Dynamics}\label{sec:03}

Next we derive the reduced macroscopic dynamics of the system. Taking the thermodynamic limit $N\to\infty$ in such a way that the relative sizes of the communities for each $\sigma$, $N_\sigma/N$, remain constant, we introduce $C$ density functions $f^\sigma(\theta,\omega,t)$ that describe, for community $\sigma$ at time $t$, the fraction of oscillators with phase and frequency in the infinitesimal intervals $[\theta,\theta+d\theta)$ and $[\omega,\omega+d\omega)$, respectively. First,  due to conservation of oscillators each $f^\sigma$ must satisfy the continuity equation
\begin{align}
\frac{\partial f^\sigma}{\partial t}+\frac{\partial}{\partial\theta}\left(f^\sigma\dot{\theta}\right)=0.\label{eq:05}
\end{align}
Second, since the frequency distribution $g^\sigma(\omega)$ is fixed the Fourier series for $f^\sigma$ must take the form 
\begin{align}
f^\sigma(\theta,\omega,t)=\frac{g^\sigma(\omega)}{2\pi}\left(1+\sum_{n=1}^\infty \hat{f}_n^\sigma(\omega,t)e^{in\theta}+\text{c.c.}\right),\label{eq:06}
\end{align}
where c.c. indicates the complex conjugate of the preceding term. Following the work of Ott and Antonsen, we propose the ansatz where the Fourier coefficients decay geometrically, i.e., there is some function $\alpha^\sigma(\omega,t)$ for which $\hat{f}_n^\sigma(\omega,t)=[\alpha^\sigma(\omega,t)]^n$. Remarkably, using this ansatz in Eq.~(\ref{eq:06}) and inserting it into the continuity equation, Eq.~(\ref{eq:05}), all modes collapse onto a single differential equation for $\alpha^\sigma$, namely
\begin{align}
\frac{\partial}{\partial t}\alpha^\sigma=-i\omega\alpha^\sigma+\frac{1}{2}\left(H^*-H\alpha^{\sigma*2}\right).\label{eq:07}
\end{align}

To connect the dynamics of $\alpha^\sigma$ with those of the order parameter $z^\sigma$, we now note that in the thermodynamic limit we may write
\begin{align}
z_q^\sigma=\int_{-\infty}^{\infty}\int_0^{2\pi} f^\sigma(\theta,\omega,t)e^{qi\theta(t)} d\theta d\omega,\label{eq:08}
\end{align}
and after integrating over $\theta$ we have that
\begin{align}
z_q^\sigma=\int_{-\infty}^{\infty} [\alpha^{\sigma*}(\omega,t)]^q g^\sigma(\omega) d\omega,\label{eq:09}
\end{align}
Finally, by assuming that the natural frequency distributions are Lorentzian, i.e., $g^\sigma(\omega,t)=\Delta^\sigma/\{\pi[(\Delta^\sigma)^2+(\omega-\omega_0^\sigma)^2]\}$ we may evaluate Eq.~(\ref{eq:09}) using the simple pole at $\omega=\omega_0^\sigma-i\Delta^\sigma$ and the Cauchy residue theorem~\cite{}, yielding
\begin{align}
z^\sigma(t)=\alpha^{\sigma*}(\omega_0^\sigma-i\Delta^\sigma,t),\label{eq:10}
\end{align}
and
\begin{align}
z_2^\sigma(t)=[z^\sigma(t)]^2\label{eq:11}
\end{align}
Thus, by evaluating Eq.~(\ref{eq:07}) at $\omega=\omega_0^\sigma-i\Delta^\sigma$ and taking a complex conjugate, we obtain the following closed system for the $C$ order parameters:
\begin{widetext}
\begin{align}
\dot{z}^\sigma=i\omega_0^\sigma z^\sigma-\Delta^\sigma z^\sigma +& \frac{1}{2}\left\{\left(\sum_{\sigma'=1}^C K_1^{\sigma\sigma'}z^{\sigma'}\right) + \left(\sum_{\sigma'=1}^C\sum_{\sigma''=1}^C K_2^{\sigma\sigma'\sigma''}(z^{\sigma'})^2z^{\sigma''*}\right)\right.\nonumber\\
&\left.-\left[\left(\sum_{\sigma'=1}^C K_1^{\sigma\sigma'}z^{\sigma'*}\right) + \left(\sum_{\sigma'=1}^C\sum_{\sigma''=1}^C K_2^{\sigma\sigma'\sigma''}(z^{\sigma'*})^2z^{\sigma''}\right)\right](z^\sigma)^2\right\}.\label{eq:12}
\end{align}
Lastly, it is convenient to treat the macroscopic dynamics described by Eq.~(\ref{eq:12}) in polar coordinates, which are given by
\begin{align}
\dot{r}^\sigma&=-\Delta^\sigma r^\sigma + \frac{1-(r^\sigma)^2}{2}\left[\sum_{\sigma'=1}^CK_1^{\sigma\sigma'}r^{\sigma'}\cos(\psi^{\sigma'}-\psi^\sigma)+\sum_{\sigma'=1}^C\sum_{\sigma''=1}^CK_2^{\sigma\sigma'\sigma''}(r^{\sigma'})^2r^{\sigma'}\cos(2\psi^{\sigma'}-\psi^{\sigma''}-\psi^\sigma)\right],\label{eq:13}\\
\dot{\psi}^\sigma&=\omega_0^\sigma + \frac{1+(r^\sigma)^2}{2r^\sigma}\left[\sum_{\sigma'=1}^CK_1^{\sigma\sigma'}r^{\sigma'}\sin(\psi^{\sigma'}-\psi^\sigma)+\sum_{\sigma'=1}^C\sum_{\sigma''=1}^CK_2^{\sigma\sigma'\sigma''}(r^{\sigma'})^2r^{\sigma'}\sin(2\psi^{\sigma'}-\psi^{\sigma''}-\psi^\sigma)\right].\label{eq:14}
\end{align}
\end{widetext}

Before proceeding to a more in-depth analysis of the reduced equations given in Eqs.~(\ref{eq:13}) and (\ref{eq:14}), we discuss more precisely the nature of coupling strengths in light of community structure and the effect on higher-order interactions. The implication of community structure on dyadic interactions is relatively straightforward, namely, coupling between pairs of oscillators within the same community must be stronger than coupling between oscillators in different communities, i.e., for $\sigma\ne\sigma'$ we have that
\begin{align}
K_1^{\sigma\sigma}\ge K_1^{\sigma\sigma'},\label{eq:15}
\end{align}
where equality is only obtained in the limit where community structure vanishes. On the other hand, the implication of community structure in the context of  triadic coupling is more complicated. Here we will assume that the effects of community structure on triadic coupling act in such a way that the triadic coupling strengths generically decrease as the oscillators belong to different communities, i.e., for $\sigma\ne\sigma'$, $\sigma\ne\sigma''$, and $\sigma'\ne\sigma''$ we have that
\begin{align}
K_2^{\sigma\sigma\sigma}\ge K_2^{\sigma\sigma\sigma'}=K_2^{\sigma\sigma'\sigma}\ge K_2^{\sigma\sigma'\sigma'}\ge K_2^{\sigma\sigma'\sigma''}.\label{eq:16}
\end{align}

\begin{figure}[b]
\centering
\epsfig{file =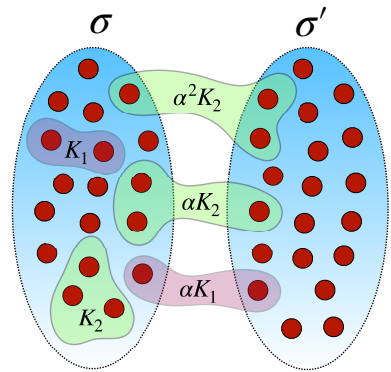, clip =,width=0.6\linewidth }
\caption{{\it Community structure and higher-order coupling.} In the case of two communities, the coupling strength associated to an oscillator in community $\sigma$ via a hyperedge of size $m$ is given by $\alpha^j K_{m-1}$, where $j$ is the number of oscillators in the hyperedge {\it not} belonging to community $\sigma$.}\label{cartoon}
\end{figure}

Note that the left- and right-most terms represent the triadic coupling strength between oscillators that are, respectively, all in the same community and all in different communities. On the other hand, the three middle terms represent triadic coupling between a triplet of oscillators that collectively belong to two communities. We take these to be all equal except for $K_2^{\sigma\sigma'\sigma'}$, which we assume is small since the affected oscillator belongs to a different community than the other two. To capture this hierarchy of coupling in a simple way we introduce a parameter $\alpha\in[-1,1]$ that scales coupling strengths, namely for $\sigma\ne\sigma'$, $\sigma\ne\sigma''$, and $\sigma'\ne\sigma''$, dyadic coupling strengths are given by
\begin{align}
K_1^{\sigma\sigma}=K_1,~~~\text{and}~~~K_1^{\sigma\sigma'}=\alpha K_1,\label{eq:17}
\end{align}
and triadic coupling strengths are given by
\begin{align}
&K_2^{\sigma\sigma\sigma}=K_2,~~~K_2^{\sigma\sigma\sigma'}=K_2^{\sigma\sigma'\sigma}=\alpha K_2,\nonumber\\
&K_2^{\sigma\sigma'\sigma'}=\alpha^2 K_2,~~~\text{and}~~~K_2^{\sigma\sigma'\sigma''}=\alpha^3 K_2.\label{eq:18}
\end{align}
This coupling structure is illustrated schematically for the case of two communities in Fig.~\ref{cartoon}. Note that in such a case with only two communities, which will be our focus for the remainder of this paper, the weakest form of triadic coupling plays no role.

\begin{figure}[t]
\centering
\epsfig{file =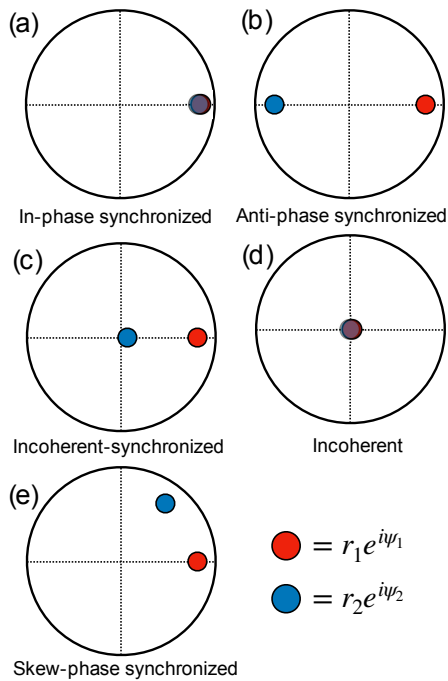, clip =,width=0.7\linewidth }
\caption{{\it Stable states.} Schematic illustration of the five possible stable states in the two community case: (a) in-phase synchronized, (b) anti-phase synchronized, (c) incoherent-synchronized, (d) incoherent, and (e) skew-phase synchronized.}\label{fig2}
\end{figure}

\section{Two communities}\label{sec:04}

We now turn to present a detailed analysis of the two community case, for which the macroscopic dynamics are described by the amplitudes, $r_1$ and $r_2$, and phases, $\psi_1$ and $\psi_2$, of the respective communities. In fact, the dynamics can be further reduced to three dimensions by introducing the phase difference $\phi = \psi_1-\psi_2$, which yields the following system of equations
\begin{widetext}
\begin{align}
\dot{r}_1&=-\Delta_1 r_1 + \frac{1-r_1^2}{2}\left[K_1r_1 + \alpha K_1r_2\cos(\phi) + K_2r_1^3 + \alpha K_2r_1^2r_2\cos(\phi) + \alpha K_2r_2^2r_1\cos(2\phi)+\alpha^2K_2r_2^3\cos(\phi)\right],\label{eq:19}\\
\dot{r}_2&=-\Delta_2 r_2 + \frac{1-r_2^2}{2}\left[\alpha K_1r_1 \cos(\phi)+ K_1r_2 + \alpha^2K_2r_1^3\cos(\phi) + \alpha K_2r_1^2r_2\cos(2\phi) + \alpha K_2r_2^2r_1\cos(\phi)+K_2r_2^3\right],\label{eq:20}\\
\dot{\phi}&=\delta\omega - \frac{\alpha}{2}\left[K_1\left(\frac{r_2^2(1+r_1^2)+r_1^2(1+r_2^2)}{r_1r_2}\right)\sin(\phi) + K_2\left(\frac{r_1^2r_2^2(1+r_1^2)+r_1^2r_2^2(1+r_2^2)}{r_1r_2}\right)\sin(-\phi)\right.\nonumber\\
&~~~~~~~~~~~~~~~~~~~\left.+ K_2\left(\frac{r_2^3r_1(1+r_1^2)+r_1^3r_2(1+r_2^2)}{r_1r_2}\right)\sin(2\phi) + \alpha K_2\left(\frac{r_2^4(1+r_1^2)+r_1^4(1+r_2^2)}{r_1r_2}\right)\sin(\phi)\right],\label{eq:21}
\end{align}
\end{widetext}
where $\delta\omega=\omega_0^{1}-\omega_0^{2}$. Since our focus is on the effect of community structure, we assume that the aggregate local dynamics of each community is the same, so that $\Delta_1=\Delta_2=\Delta$ and $\omega_0^1=\omega_0^2$, yielding $\delta\omega$=0. In this case we find five different qualitatively-different steady-states, depending on the choice of parameters and initial conditions:
\begin{enumerate}[(i)]
\item An {\it in-phase synchronized} state with $r_1 = r_2 > 0$ and $\phi = 0$.
\item An {\it anti-phase synchronized} state with $r_1 = r_2 > 0$ and $\phi = \pi$.
\item An {\it incoherent-synchronized} state with $r_1 > 0$, $r_2 \approx 0$ and $\phi = 0$.
\item An {\it incoherent} state with $r_1 = r_2 = 0$.
\item A {\it skew-phase synchronized} state with $r_1 = r_2 > 0$ and $0<\phi<\pi/2$.
\end{enumerate}
These five states are illustrated in Fig.~\ref{fig2}(a)--(e), respectively, with red and blue circles denoting the locations of the order parameters $r_1e^{\psi_1}$ and $r_2e^{i\psi_2}$ in each case, and purple circles denoting overlapping order parameters. As we will see below, the four states (a)--(d) are all supported by the system dynamics when interaction between communities is cooperative, i.e., $\alpha>0$. On the other hand, we find that when interaction between communities is contrarian, i.e., $\alpha<0$, the incoherent and skew-phase synchronized states are supported.

\begin{figure}[t]
\centering
\epsfig{file =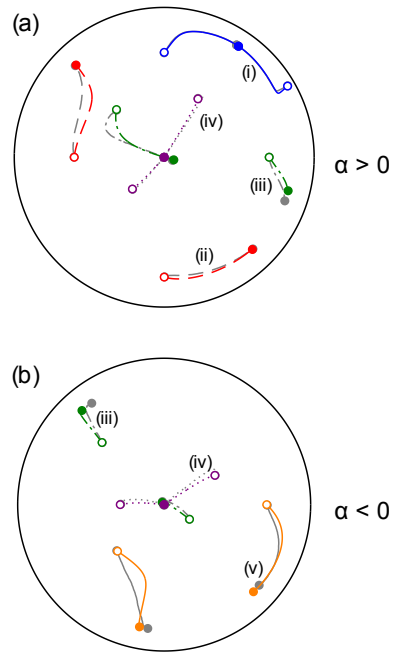, clip =,width=0.65\linewidth }
\caption{{\it Multistability in systems with higher-order interactions and community structure.} Illustration of the possible states using trajectories of order parameters on the complex unit disc for (a) positive and (b) negative $\alpha$. Parameters are given by $\Delta = 1$, $K_1=0.1$, $K_2=10$, and $\alpha = 0.1$ or $-0.1$. The labels (i), (ii), (iii), (iv), and (v) correspond to the in-phase synchronized state, the anti-phase synchronized state, the incoherent-synchronized state, the incoherent state, and the skew-phase synchronized state. Gray curves describe the dynamics of direct simulations with $N=10^5$ total oscillators.}\label{fig3}
\end{figure}

Before moving to our analytical results, we illustrate the dynamics and the multistability between these different states with some numerical simulations. Finding the richest dynamics occurring for a combination of small dyadic coupling and large triadic coupling, in Fig.~\ref{fig3}(a) and (b) we plot the trajectories of the dynamics of the two order parameters in the complex unit disc for (a) four and (b) two different initial conditions using $\Delta = 1$, $K_1=0.1$, $K_2=10$, and (a) $\alpha = 0.1$ and (b) $-0.1$. Initial and final states are plotted as open and filled circles, respectively, and trajectories that end up at the different states are labelled: (i) in-phase synchronized, (ii) anti-phase synchronized, (iii) incoherent-synchronized, (iv) incoherent, and (v) skew-phase synchronized. These trajectories are plotted in solid blue, dashed red, dot-dashed green, dotted purple, and solid orange, respectively. Note that for these parameters all four states (i)--(iv) are stable for $\alpha=0.1$ and both states (iv) and (v) are stable for $\alpha=-0.1$. Grey lines in Fig.~\ref{fig3}(a) correspond to direct simulations of the full system (\ref{eq:01}) with a total of $N=10^5$ oscillators, i.e., $N_\sigma=5\times10^4$ in each community, which show good agreement with the reduced dynamics in Eqs.~(\ref{eq:13})-(\ref{eq:14}).




Moving forward, we aim to understand these states and their dynamics using analytical methods. To this end, we note that the collection of five possible states lies on two manifolds within the full phase space. The first of these we call the {\it $r$-manifold}, and is characterized by $r_1=r_2=r$, i.e., both communities are characterized by exactly the same degree of synchronization. The second manifold, which we call the {\it $\phi$-manifold}, is characterized by $\phi=0$, i.e., the two communities share the same mean phase. Note that states (i), (ii), (iv), and (v) lie on the $r$-manifold, while states (i), (iii), and (iv) lie on the $\phi$-manifold. We will now focus on these two manifolds to give analytical insight into the system dynamics.

\begin{figure}[t] 
\centering
\epsfig{file =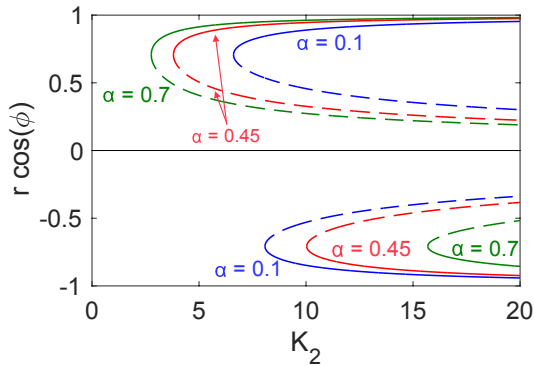, clip =,width=0.9\linewidth }
\caption{{\it Stable and unstable branches: in-phase and anti-phase synchronized states.} Fixed points $r_a^{+}$, $r_b^{+}$ (solid lines) and $r_a^{-}$, $r_b^{-}$ (dashed lines) as a function of $K_2$ for $K_1=0$, $\Delta = 1$, and $\alpha = 0.1$, $0.45$, and $0.7$.}\label{fig4}
\end{figure}

\subsection{The $r$-manifold}\label{subsec:04:01}

Beginning with the case $r_1=r_2=r$, we also make the simplifying assumption that $K_1=0$, i.e., coupling is solely triadic. As we will see, this suffices to recover all five steady-states illustrated above. On this manifold the dynamics then reduce to
\begin{align}
\dot r &= -\Delta r + \frac{K_2 r^3(1-r^2)}{2}\nonumber\\
&\times\left[ 1+\alpha \cos(\phi) + \alpha \cos(2\phi) \alpha^2 \cos(\phi)\right],\label{eq:22}\\
\dot \phi &= -K_2 r^2 \alpha (1+r^2)\left[ -\sin(\phi) +\sin(2\phi) + \alpha \sin(\phi)\right],\label{eq:23}
\end{align}
for which, in addition to the incoherent state $r=0$, we find fixed points $(r_a^-,0)$, $(r_a^+,0)$, $(r_b^-,\pi)$, $(r_b^+,\pi)$, $(r_b^-,\pm  \phi^{\text{skew}})$, and $(r_b^+,\pm \phi^{\text{skew}})$, where
\begin{align}
r_a^\pm = \sqrt{\frac{1}{2} \pm \frac{\sqrt{(\alpha +1)^2 K_2-8}}{2 (\alpha +1) \sqrt{K_2}}},\label{eq:24}\\
r_b^\pm = \sqrt{\frac{1}{2} \pm \frac{\sqrt{\left(\alpha ^2-1\right) K_2+8}}{2 \sqrt{\alpha ^2-1} \sqrt{K_2}}},\label{eq:25}\\
\phi^{\text{skew}} = \pm \arccos\left(\frac{1-\alpha}{2}\right). \label{eq:26}
\end{align}
Note first that when $\alpha = 0$ (i.e., the communities are isolated from one another), the branches $r_a^{\pm}$ and $r_b^{\pm}$ are equivalent. Next, when $0< \alpha < 1$ the branches $r_a^{\pm}$ and $r_b^{\pm}$ appear at saddle-node bifurcations, respectively, at 
\begin{align}
K_2^a = \frac{8}{(1+\alpha)^2} ~~~~~~~~\text{and}~~~~~~~~ K_2^b = \frac{8}{1-\alpha^2}.\label{eq:27}
\end{align}
After the saddle-node bifurcation occurs, the branches $r_a^+$ and $r_b^+$ are stable fixed points that correspond to the in-phase synchronized and anti-phase synchronized states, respectively, with $r_a^-$ and $r_b^-$ being their unstable counterparts. (By linearization, see Appendix~\ref{appendixA}, it can be shown that the  $r_a^+$ and $r_b^+$ are stable and $r_a^-$ and $r_b^-$ are unstable.) These branches are illustrated in Fig.~\ref{fig4}, depicting $r\cos(\phi)$ as a function of $K_2$ for $\alpha=0.1$, $0.45$, and $0.7$ (blue, red, and green, respectively). Note that the two branches occur at similar values of $K_2$ for small $\alpha$, i.e., strong community structure, and as $\alpha$ increases, i.e., community structure becomes weaker, the branches occur at $K_2$ values that become more different, as larger $\alpha$ promotes in-phase synchronization but pushes back the anti-phase synchronized branch.

\begin{figure}[t] 
\centering
\epsfig{file =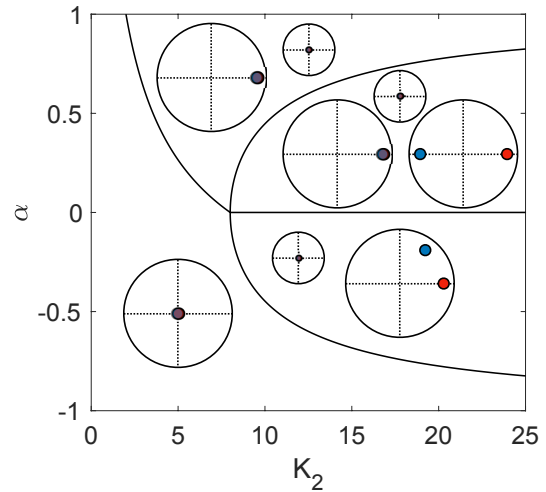, clip =,width=0.95\linewidth }
\caption{{\it Stability diagram for the $r$-manifold.} Diagram illustrating the stability of different states in the $r$-manifold for $K_1=0$ and $\Delta = 1$. }\label{fig5}
\end{figure}

Lastly, for $-1<\alpha<0$ the system displays different synchronized behavior. Rather than supporting the in-phase and anti-phase synchronized states, these become unstable and the skew-phase synchronized state becomes stable, taking the same local degrees of synchronization as the anti-phase synchronized state, i.e., $r=r_b^{\pm}$ (again, with $r_b^+$ being the stable fixed point), but interestingly, the two organize into two clusters that lie at a phase difference $\phi^{\text{skew}}\in(0,\pi/2)$. The system transitions directly from the incoherent state to a stable state where both communities are synchronized with the same value of the order parameter $r_b^+$, but have a phase difference $0 < \phi^{\text{off}} < \pi$ [see Eq.~(\ref{eq:26})]. 

The stability diagram for the $r$-manifold for the case of $K_1 = 0$ is summarized in Fig.~\ref{fig5}, which shows the stable steady states as a function of the parameters $K_2$ and $\alpha$. For positive values of $\alpha$, as $K_2$ is increased, the incoherent state is first the only stable state, after which the in-phase synchronized state is born at $K_2^a$ in a saddle-node bifurcation, and subsequently the anti-phase synchronized state is born at $K_2^b$, again in a saddle-node bifurcation. On the other hand, for negative $\alpha$, as $K_2$ increases, the incoherent state is, again, first the only stable state, after which only one saddle-node bifurcation occurs at $K_2^b$, giving rise to the skew-phase synchronized state.

\subsection{The $\phi$-manifold}\label{subsec:04:02}

Next, we proceed to analyze the $\phi$-manifold, specifically with the aim of gaining some understanding of the incoherent-synchronized state, which is the only observed steady-state not on the $r$-manifold analyzed above. To this end, focusing again on the case of $K_1=0$, we assume that $\phi=0$, yielding the reduced set of equations
\begin{align}
\dot{r}_1&=-\Delta r_1 + \frac{1-r_1^2}{2}\nonumber\\&\times(K_2r_1^3 + \alpha K_2r_1^2r_2 + \alpha K_2r_2^2r_1+\alpha^2K_2r_2^3),\label{eq:28}\\
\dot{r}_2&=-\Delta r_2 + \frac{1-r_2^2}{2}\nonumber\\&\times(\alpha^2K_2r_1^3 + \alpha K_2r_1^2r_2 + \alpha K_2r_2^2r_1+K_2r_2^3).\label{eq:29}
\end{align}
While an exact analytical expression for the incoherent-synchronized state is difficult to obtain, an approximation is tractable using a perturbation analysis. First, we note that, provided that $K_2\ge8$, in the limit $\alpha\to0$ (i.e., when community structure is so strong that the two communities are isolated from one another), both $r_{1,2}=0$ and $\sqrt{\frac{1}{2}+\frac{\sqrt{K_2-8\Delta}}{2\sqrt{K_2}}}$ are stable. Moreover, the incoherent-synchronized state is only stable for sufficiently strong community structure, thereby making $\alpha$ a suitable perturbation parameter. We then consider a perturbation of the steady-state $r_1>0$, $r_2=0$ by the parameter $\alpha$, i.e., searching for solutions with expansions
\begin{align}
r_1=r_1^{(0)}+ r_1^{(1)}\alpha+r_1^{(2)}\alpha^2 + r_1^{(3)}\alpha^3+\mathcal{O}(\alpha^4),\label{eq:30}\\
r_2=r_2^{(0)}+ r_2^{(1)}\alpha+r_2^{(2)}\alpha^2 + r_2^{(3)}\alpha^3+\mathcal{O}(\alpha^4).\label{eq:31}
\end{align}
Inserting Eqs.~(\ref{eq:30}) and (\ref{eq:31}) into Eqs.~(\ref{eq:28}) and (\ref{eq:29}) and setting $\dot{r}_1,\dot{r}_2=0$ yields
\begin{widetext}
\begin{align}
2\Delta \left(r_1^{(0)} +r_1^{(1)}\alpha + r_1^{(2)}\alpha^2 + r_1^{(3)}\alpha^3 \right)&= \left[K_2r_1^{(0)3} (1-r_1^{(0)2}) \right]\nonumber\\
&+\left[r_1^{(1)} (3 K_2 r_1^{(0)2} - 5 K_2 r_1^{(0)4})\right]\alpha\nonumber\\
 &+ \left[  K_2 r_1^{(0)} \left((3 - 10 r_1^{(0)2}) r_1^{(1)2} + r_1^{(0)} (3 r_1^{(2)} - 5 r_1^{(0)2} r_1^{(2)} + r_2^{(1)} - r_1^{(0)2} r_2^{(1)})\right)\right]\alpha^2\nonumber\\
 &+\left[K_2 ((1 - 10 r_1^{(0)2}) r_1^{(1)3} + 2 r_1^{(0)} r_1^{(1)} ((3 - 10 r_1^{(0)2}) r_1^{(2)} + r_2^{(1)} - 2 r_1^{(0)2} r_2^{(1)}) \right.\nonumber\\
 &~~~~+ \left.r_1^{(0)} (r_2^{(1)2} + r_1^{(0)} ((3 - 5 r_1^{(0)2}) r_1^{(3)} + r_2^{(2)} - r_1^{(0)} (r_2^{(1)2} + r_1^{(0)} r_2^{(2)}))))\right]\alpha^3,\label{eq:32}\\
 2\Delta\left(r_2^{(1)}\alpha + r_2^{(2)}\alpha^2 + r_2^{(3)}\alpha^3\right)&=\left[ K_2 r_1^{(0)2} (r_1^{(0)} + r_2^{(1)})\right]\alpha^2\nonumber\\
 &+\left[K_2 (r_2^{(1)3} +  r_1^{(0)} r_2^{(1)} (2 r_1^{(1)} + r_2^{(1)}) + r_1^{(0)2} (3 r_1^{(1)} + r_2^{(2)}))\right]\alpha^3,\label{eq:33}
\end{align}
where we've used that $r_2^{(0)}=0$. Collecting terms at different orders of $\alpha$, we solve to obtain
\begin{align}
r_1^{(0)}&=\sqrt{\frac{1}{2}+\frac{\sqrt{K_2-8\Delta}}{2\sqrt{K_2}}},\label{eq:34}\\
r_1^{(1)}&=0,\label{eq:35}\\
r_1^{(2)}&=-\frac{K_2r_1^{(0)2}(1-r_1^{(0)2})r_2^{(1)}}{2\Delta + K_2r_1^{(0)2}(3-5r_1^{(0)2})}=0,\label{eq:36}\\
r_1^{(3)}&=-\frac{K_2r_1^{(0)}(1-r_1^{(0)2})(r_2^{(1)2}+r_1^{(0)}r_2^{(2)})}{2\Delta + K_2r_1^{(0)2}(3-5r_1^{(0)2})}=\frac{\sqrt{K_2}}{2\sqrt{2}\sqrt{K_2-8\Delta}}\sqrt{1+\frac{\sqrt{K_2-8\Delta}}{\sqrt{K_2}}}.\label{eq:37}\\
r_2^{(0)}&=0,\label{eq:38}\\
r_2^{(1)}&=0,\label{eq:39}\\
r_2^{(2)}&=\frac{K_2r_1^{(0)2}(r_1^{(0)}+r_2^{(1)})}{2\Delta}=\frac{K_2}{4\sqrt{2}\Delta}\left(1+\frac{\sqrt{K_2-8\Delta}}{\sqrt{K_2}}\right)^{3/2},\label{eq:40}\\
r_2^{(3)}&=\frac{ K_2 (r_2^{(1)2} (r_1^{(0)} + r_2^{(1)}) + r_1^{(0)2} r_2^{(2)})}{2\Delta}=\frac{K_2^2}{16\sqrt{2}\Delta^2}\left(1+\frac{\sqrt{K_2-8\Delta}}{\sqrt{K_2}}\right)^{5/2}.\label{eq:41}
\end{align}
\end{widetext}

\begin{figure}[t] 
\centering
\epsfig{file =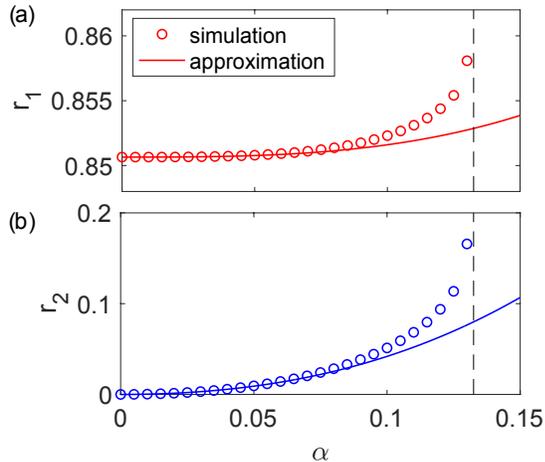, clip =,width=0.9\linewidth }
\caption{{\it Incoherent-synchronized state: perturbation theory.} Comparison of the perturbation theory from the text (solid curves) to direct simulation of the low dimensional dynamics for the incoherent-synchronized state using $K_1=0$, $K_2=10$, and $\Delta = 1$. }\label{fig6}
\end{figure}

Note that, with $r_1^{(1)}$, $r_1^{(2)}$, and $r_2^{(1)}$ all zero, the leading order correction to the order parameters $r_1$ and $r_2$ for the synchronized and incoherent come at cubic and quadratic orders in $\alpha$, respectively. In Fig.~\ref{fig6} we compare the perturbation theory obtained above (solid curves) to direct simulations of the low dimensional dynamics (circles) using $K_2=10$ and $\Delta=1$ for both $r_1$ and $r_2$ in panels (a) and (b). We note strong agreements until near the critical value of $\alpha$ where the incoherent-synchronized state disappears, denoted by the vertical dashed lines.

\section{Discussion}\label{sec:05}

We have studied the synchronization of coupled phase oscillators with community structure and higher-order interactions. When there is no community structure present, as in Refs.~\cite{Skardal2020CommPhys}, higher-order interactions result in bistable states where both the incoherent and synchronized states are stable. Therefore, one would naturally expect that for sufficiently strong community structure there would be stable states where both communities are synchronized, both are incoherent, and one is synchronized while the other is incoherent. While these states are, indeed, observed for sufficiently strong community structure, we have found that two additional states can appear. The first is a state where both communities are synchronized but are clustered opposite one another. This anti-phase synchronized state requires {\it both} strong enough community structure and higher-order coupling strength (see Fig.~\ref{fig5}). Since community structure and higher-order interactions are observed in brain oscillations, we hypothesize that this mechanism could be relevant for producing anti-phase oscillations in brain rhythms. In addition, we found an additional novel state where both communities are synchronized but oscillate with a phase difference that can be tuned by the amount of community structure [see Eq.~(\ref{eq:26})]. This skew-phase state is observed only when the sign of coupling via triadic interactions alternates depending on the membership of oscillators in the triad to different communities.

In order to focus on the more novel results, we studied particular cases of the full model in Eq.~(\ref{eq:01}). The most important simplifications were that we considered only triadic interactions, two communities, and that we assumed that the two communities had identical frequency distributions ($\Delta_1=\Delta_2$ and $\delta \omega = 0$). Here we hypothesize on what should be expected if these assumptions are removed, but leave a detailed study for future work. The addition of a nonzero mean frequency difference between communities ($\delta \omega > 0$) should produce bifurcations to standing wave solutions with two counterrotating groups of oscillators (as in, e.g., \cite{Martens2009PRE}). The hierarchical coupling structure in Eqs.~(\ref{eq:18}) could be generalized in a straightforward manner to cases with larger interactions and more communities, and the subsequent analysis would follow in a similar way. The presence of more communities should result in stable states where some communities are synchronized while the rest are almost incoherent. How higher-order interactions and community structure would affect the relative phases of the synchronized communities or if additional stable states could be present is not clear, however, and is left for future study. In summary, by analyzing the simplest phase oscillator model that includes community structure and higher-order interactions, we have found that multiple synchronized states can coexist, including anti-phase and skew-phase synchronized states. We anticipate that the interplay of community structure and higher-order interactions in less simplified scenarios will result in complex oscillation dynamics.

\appendix

\section{Stability of the in-phase and anti-phase states}\label{appendixA}

In this Appendix we provide detailed calculations for the linear stability of the in-phase and anti-phase states discussed in Sec.~\ref{sec:04}. We start by calculating the eigenvalues Jacobian associated to Eqs.~(\ref{eq:19})-(\ref{eq:21}) with $K_1  = 0$ at the in-phase and anti-phase fixed points. 

At $r_1 = r_2 = r_a^+$, $\phi = 0$, the eigenvalues of the Jacobian are, letting $k \equiv K_2/K_2^a$,
\begin{align}
\lambda_1&= \frac{2\alpha }{\alpha +1}\left(1 - 4k\right)  -\alpha  \sqrt{8 K_2} \sqrt{k-1},\\
\lambda_2 &=\frac{4-2 \alpha}{(\alpha +1)} -4 k-(\alpha + 1)\sqrt{2K_2}\sqrt{k-1},\\
\lambda_3 &= 4\left(1-k\right)-\frac{1}{2} (\alpha +1) \sqrt{8 K_2} \sqrt{k-1},
\end{align}
which are negative for $ \alpha > 0$ and $K_2 > K_2^a$ ($k > 1$), confirming the linear stability of the fixed point $r_1 = r_2 = r_a^+$, $\phi = 0$ . 

At $r_1 = r_2 = r_a^-$, $\phi = 0$, the eigenvalues are
\begin{align}
\lambda_1&=\frac{2 \alpha  \left(-4 k+4 \sqrt{k-1} \sqrt{k}+1\right)}{\alpha +1},\\
\lambda_2 &=\frac{6}{\alpha +1}-4 k+4 \sqrt{k-1} \sqrt{k}-2,\\
   \lambda_3 &= 4(1-k)+4 \sqrt{k-1} \sqrt{k}.
\end{align}
For $k > 1$ the eigenvalue $\lambda_3$ is positive, and therefore the fixed point $r_1 = r_2 = r_a^-$, $\phi = 0$ is unstable. The linear stability and instability of the fixed points $r_1 = r_2 = r_b^+$, $\phi = \pi$ and $r_1 = r_2 = r_a^-$, $\phi = \pi$, respectively, can be checked similarly.


\acknowledgements
PSS acknowledges NSF grant MCB-2126177.

\bibliographystyle{unsrt}

\end{document}